\documentstyle[12pt]{article}
\begin{document}

\title{
Evolution to the Edge of Chaos in Imitation Game
}
\author{
        Kunihiko Kaneko and Junji Suzuki\\
        {\small \sl Department of Pure and Applied Sciences,}\\
        {\small College of Arts and Sciences}\\
        {\small \sl University of Tokyo, Komaba 3-8-1, Meguro-ku, Tokyo 153}
 \\}
\date{}
\maketitle
\begin{abstract}
Motivated by the evolution of complex bird songs, an abstract
imitation game is proposed to study the increase of dynamical complexity:
Artificial "birds" display a "song" time series to each other,
and those imitate the other's song better win the game.  With the
introduction of population dynamics according to the score of the game
and the mutation of parameters for the
song dynamics, the dynamics is found to evolve towards the borderline
between chaos and a periodic window, after punctuated equilibria.
The importance of edge of chaos with topological chaos for complexity is
stressed.
\end{abstract}


\vspace{.3in}

\section{Introduction}

Increase of complexity through evolution is believed to be seen in
many biological systems, not only in the hierarchical
organization in genotypes and phenotypes but also in animal behavior and
communication.  A bird song, for example, increases its repertoire through
evolution and developments \cite{Catch}. In the bird song, two functions
are believed to exist;
one is the defense of territory, and the other is the sexual attraction.
It is known that a bird with a complex song ( with many repertoires
made from combinations of simple phrases) is stronger in defending its
territory, as is demonstrated by Krebs with the help of loud-speaker
experiments \cite{Catch,Krebs}.


There are some reports of observations that birds try to
imitate each other's song at the defense of territory \cite{Catch}.
Inspired by these observations, we propose an
imitation game of birds with their songs, for the defense of territory.
A complex song may not easily be imitated, and may be powerful in the defense
of territory.
Although we do not claim the plausibility of this hypothesis strongly
here, the evolution in an imitation game is of interest of itself,
as a novel evolution game \cite{Maynard} and
in a more general context for the evolution of communication
or mimicry \cite{Maynard2}. Here we consider an abstract game for the
imitation, where an artificial "bird" wins the game when it can imitate the
other's song better than the other player bird.  If a song is
simple, it may easily be imitated by others, and we may expect the
emergence of evolution towards a complex song.

The model we propose here has sound basis on nonlinear dynamics.
It adopts a nonlinear mapping as the generator of songs.
Our birds thus can produce songs possessing the complexity of real
numbers, which is one of the advantages of our model.

Another motivation of the present paper is the introduction of
a simple model which realizes the "evolution to the
edge of chaos".  The increase of complexity at the edge of chaos
has been discussed in cellular automata
\cite{Chris}, in a Boolean network \cite{Kauf2},
and in coupled map lattices \cite{KK}.  In a system with mutually
interacting units,
it has been suggested that this edge of chaos has potential
advantages in evolution \cite{Packard}.
Although these studies are reasonable in their own contexts,
there is no clear simple
example which provides an evolution to the edge of chaos, in
the exact sense of dynamical systems theory ( note that chaos
is defined only on dynamical systems with a continuos state,
and is not defined for a discrete-state system like cellular automata).
Since we have adopted a nonlinear dynamical system as a song generator and an
imitator, one can examine if a song evolves towards the edge of chaos,
which is a clear advantage of our model.

Indeed, we will find that the dynamics of the song time
series evolves towards the edge of chaos through the imitation game here.
This edge state, however, lies not at the onset of chaos, but at the edge
between a window of a periodic cycle and
chaos.  The main difference here is that chaotic orbits exist as a transient
orbit in the windows.  The importance of the existence of transient chaos will
also be discussed.

\section{Modeling}

Our abstract model consists of the following processes.

\vspace{.1in}
{\bf a: Song dynamics}:
As a "song", we use a time series generated by a simple
mapping $x_{n+1}=1-ax_n ^2$, the logistic map \cite{logistic0}.
The attractor of
the map shows a bifurcation sequence from a
fixed point, to cycles with periods 2,4,8,..., and
to chaos as the parameter $a$ is increased \cite{logistic}.
The parameter value $a$ is assumed to be
different by each individual "bird".  We choose this dynamics since it
is investigated in detail as the simplest generic model for bifurcation and
chaos.

\vspace{.1in}
{\bf b: Imitation process}:
Each bird player $i$ chooses an initial condition, so that
the time series by its own dynamics $x_{n+1}(i)=f_i (x)=1-a(i) x_n (i)^2$ can
imitate the
song of the other player.  Here we use the following imitation
process for simplicity.
For given transient time steps $T_{imi}$, a player 1 modifies its dynamics
with a feedback from the other player:

\begin{equation}
x_{n+1}(1)=f_1 [(1-\epsilon)x_n (1)+\epsilon x_n (2)]
\end{equation}

By this dynamics, the player 1 adjusts its value $x_n(1)$ by referring to the
other player's value ($x_n(2)$).  Here $\epsilon $ is a coupling parameter
for imitation process.  After repeating this imitation process
for $T_{imi}$ steps, the player 1 uses its own dynamics
$x_{n+1}(1)=f_1 (x_n(1))$.  In other words, the above process is used
as a choice of initial condition for the imitation of the
other player's dynamics.  The coupling parameter $\epsilon $ also varies by
players. (However, the distribution of the parameter $\epsilon $
seems "irrelevant", judging from simulations.  We will skip the discussion on
this parameter in later sections.)

\vspace{.1in}
{\bf c: Game}:
We adopt a 2-person game between "birds".
After completing the imitation process, we measure the Euclidian distance
$D(1,2)=\sum_{m=1}^{T} |x_m (1)-x_m (2)|^2$ over certain
time steps $T$.  By changing the role of the player 1 and 2,
we measure $D(2,1)$.
If $D(1,2)<D(2,1)$, the player 1 imitates better than 2, thus being
the winner of the game, and vice versa.

\vspace{.1in}
{\sl  topology for the players}

Here all "birds" are assumed to play the imitation game against all others
with equal probability, although some simulations with the
use of a two-dimensional lattice with nearest neighbor's play
are briefly discussed later.

After each game, the winner gets
a point $W$, while a loser gets L ($W>L$)  (Both get $(W+L)/2$ in the
case of draw.) After iterating a large number of games, the population
is updated in proportional to the score.  This update of population
corresponds to the reproduction with the survival of the fitter.

\vspace{.1in}
{\bf d: Mutation}:

The parameters $a$ ($0<a<2$) and
$\epsilon $ ($0<\epsilon <1$) can be different by individuals, and
changed by mutation \cite{GA}. In the update stage of population,
mutational errors are introduced to the parameters:  The parameters $a$
and $\epsilon$ are changed to
$a+ \delta$, $\epsilon + \delta '$, where  variables $\delta$ and
$\delta '$ are random numbers chosen from a suitable distribution
( we use a homogeneous random distribution over $[-\mu , \mu]$ or Lorenzian
distribution ($P(\delta )= 1 /\{ \mu (1+ (\delta /\mu )^2)\} $).  The latter
choice is often useful, since it can provide a larger variety of species.
The former choice, on the other hand, inhibits a large jump of
parameters,
and often the parameter values are trapped at intermediate values.

\section{ Results of  Simulations}

First of all, we start from a game without population dynamics.
The scoretable of the game is plotted in Fig.1.
We note that the outcome of the game can depend on the initial choice of
$x_n(i)$ for the players. We have computed  the score of
the player 1 with the parameter $a(1)$ against the player with the
parameter $a(2)$, averaged from 100 initial values chosen randomly.
In Fig.1  the averaged score of the player with $a(1)$ against the player
$a(2)$ is plotted.

\vspace{.2in}
-------Fig.1------------

\vspace{.2in}

Although the scoretable is rather complicated, we can see some
bands of parameters strong against a wide range of parameters.
Generally these parameters lie around the bifurcation
points; bifurcation from fixed point to period 2, 2 to 4, etc.
In addition to these period doubling bifurcation points,
the borderline between chaos and windows for stable
cycles is stronger, in the high nonlinearity regime with topological
chaos ($a>1.4011...$).

With the inclusion of the population dynamics, the score of each player
can depend on the population at the moment; we do not have a fixed
fitness landscape.  To see a tendency of fitness,  we sample the scores for
each parameter range over long time steps, taking a high mutation regime,
where many players of different parameters exist.  The average score
is shown in Fig.2 with the use of bins of 0.001.
The ruggedness of landscapes is clearly seen.  The peaks of
hills are seen, for examples, at 1.25, 1.38, 1.401, 1.625, 1.77, and
1.94.  They correspond to the bifurcation points
from period-2 to 4, 4 to 8, the accumulation point of period
doubling bifurcations, the edges of windows of periods  5, 3, and 4,
respectively.

Importance of (schematic) rugged landscape is discussed in various biological
contexts.  We have to note the following points
in the present example: the fitness landscape is not given in advance,
in contrast with many theoretical models \cite{landscape}.
The rugged landscape is {\sl emergent}
through evolution.  The landscape depends strongly on the population
distribution of the moment.  Second, the figure of the landscape is
no more schematic, in contrast with other landscapes e.g., in
spin glass models \cite{SG}. Indeed, the horizontal axis of the figure 2 is
just the parameter of the logistic map, while the axis for the spinglass-type
rugged landscape is usually schematic.

\vspace{.2in}
---------Fig.2-----------
\vspace{.2in}

In Fig.3, temporal evolution of the
average of the parameter $a$ over all players is plotted.
Plateaus corresponding to the hills in Fig.2 are observed successively,
which provides an explicit example for punctuated equilibrium \cite{punceq}.
At the temporal domain with a plateau, the parameter values of "birds" are
concentrated on the plateau value.  Each plateau corresponds to the
period-doubling bifurcation or to the edge of periodic windows
among chaotic states.  For example, the plateaus in Fig.2 are
the superstable point of period-2 orbit ($a=1$; where the derivative
$f'(x_1)f'(x_2)$ changes its sign) \cite{logistic}, the
bifurcation point to period-4,
and then to period-8, and the edge of period-3 window, and finally
the edge of period-4 window.  As the mutation rate is decreased, plateaus
have longer time intervals, and many steps corresponding to
finer windows (and the accumulation point of period-doublings) are observed.

\vspace{.2in}
----------Fig.3------------
\vspace{.2in}

The evolution leads to our system to the edge of chaos, i.e.,
the borderline between chaos and a window.  To see clearly "the
edge of chaos", we have measured the  Lyapunov exponent $\lambda $.
The Lyapunov exponent is a
characteristic of asymptotic orbital instability, given by the
growth rate of tiny difference between two close orbits.  The exponent
is positive for a chaotic orbit, whose magnitude characterizes the strength
of chaos, while it is negative for a stable periodic orbit.  Thus a border
between chaos and a periodic state corresponds to $\lambda =0$.
In Fig.4 we have plotted the score of birds as a function of
the Lyapunov exponent $\lambda $, sampled for 200 time steps at
the final stage of evolution.
As is shown, the score has a broad peak around $\lambda =0$.

\vspace{.2in}
----------Fig.4----------------
\vspace{.2in}

Thus we have observed the evolution of a song to increase its complexity
in our imitation game. The dynamics of the song time series evolves
towards the edge of chaos.
Although there is no "strongest" parameter in a two-player game,
the borderlines between periodic windows and chaos are selected
in the course of the evolution.  The song at the edge ($a \approx 1.94$)
of period-4 window is strong and robust.

The robustness of the parameter depends on the mutation rate.
In the landscape (fig.2), the width of a hill
is another important quantifier besides its height.  For example, the hill
around $a\approx 1.94$ is higher than the hill around $a\approx 1.77$,
but the latter hill is much wider.  A wider hill is more robust
against the variation of the parameter, and is more advantageous in
a higher mutation rate regime.  Indeed in a simulation with a higher
mutation rate, the population is concentrated around $a\approx 1.77$.
The width of a hill is governed by the width of a window:  thus the edge of
a wider window is robust in a higher mutation rate.

We have made a large number of simulations, by changing parameters and
initial conditions.  Our conclusions are invariant against these changes
of parameters, unless the mutation rate is too small for us to observe the
successive changes within our simulation time steps (usually less than
$10^5$ steps).

\section{Modifications and Extensions}

There can be many generalizations to our model.
Here we will briefly discuss some of these, although the
results are rather preliminary as yet.

\vspace{.1in}
1) Change of the topology: Instead of the game of all to all, it may be
interesting to study a game on a two-dimensional lattice with
nearest neighbor's interaction.  Most of our results in \S 3 are
reproduced, although
spatial differentiation and clustering are found in
the lattice version for a suitable mutation rate.
The coexistence, for example, of the species near the period-3 and
period-4 windows is seen in this case.
Needless to say, "fluctuation" is much bigger in this model than that
presented in \S 3. Consequentially, some plateaus
observed previously
are lost depending on the value of mutational errors.

\vspace{.1in}
2) Symbolization of a song: Instead of judging the imitation
by the Euclidian distance, we check it after symbolizing the dynamics to
few digits ( here we choose 2).
It is well known that the time series of the logistic map can be traced
by the binary sequence (Left i.e., $x<0$ and Right i.e., $x>0$).
We check if the sign of $x_m$ is identical or not between the players 1 and 2,
at each time step and count
the number of time steps where the symbols are not matched.  In other words,
we use the
following function $I(1,2)$ instead of $D(1,2)$;
$I(1,2)=\sum_{m=1}^T sign(x_m(1),x_m(2))$ with the notation
$sign(a,b)=1$ if $ab<0$ and 0 otherwise.
Adopting this criterion, we have performed simulations of our game
on a 2D lattice.  We have found that the logistic parameter $a$ of
winners often lie at a point
where the binary sequence changes its pattern, the Right-Left sequence.
The temporal behavior of evolution here is more complicated than
that in \S 3.  Punctuated equilibrium
around $a=1.77$ can be established for some intervals;
However, this does not continue over a long time scale, and there occur
switches to different transient states.  Often species with distant
parameters can coexist.  Generally speaking, complexity and diversity
is enhanced in the present case with symbolization. This  result may give
some hint to consider the origin of
complexity and diversity in communication and language.

\vspace{.1in}
3) Dual dynamics for song and imitation:
What happens if birds assume different parameters
in singing and in imitating?
This would be a reasonable question, since there is no
primary reason that the two processes are governed by identical
dynamics.
{}From some simulations, the parameter for the singing process
again shows punctuated equilibria among "window" values,
before settling down to the border between chaos and the period-4 window value
($a \approx 1.94$). The parameter for the imitation process, on the other
hand, is often settled around $a=0.75$,
the bifurcation point from fixed point to period 2.

If we adopt the symbolization dynamics as in 2), the
dynamics is again more complicated.
The parameter for the imitation process
again shows punctuated equilibria among "window" values.
The parameter here, however, does not settle
to a window value (e.g., at $a \approx 1.94$), but wanders
upwards and downwards for ever.
On the other hand, the parameter for singing often increases up to
a value for fully
developed chaos, although there is a small amplitude variation,
in synchronization with the change of the parameter for the imitation.

\vspace{.1in}
4) Changes in the choice of mapping; This would be the most immediate idea for
most of us.  It would be far more interesting to include the evolution,
not only within the logistic maps with a variable parameter, but also
towards different types of maps.

\section{Discussions}

In the present paper we have  presented a "minimal" model showing
the evolution to the edge of chaos and punctuated equilibrium.
Indeed it gives the first explicit example for this type of evolution
in the exact sense of dynamical systems, as is
demonstrated by the evolution to a state with a Lyapunov exponent close
to zero.
In addition, we have to note that our edge state lies between a window and
chaos.  A logistic map has topological chaos at the window and can
show chaotic transients before the dynamics settles down to a
stable cycle.  The existence of transient chaos is useful to
imitate a dynamics of different nature. In the window regime,
the logistic map includes a variety of transient orbits, some of which are
close to a periodic orbit, while others are chaotic.  A window at a higher
nonlinearity regime includes a variety of unstable cycles,
 as coded by Sharkovskii's ordering
\cite{logistic}, and can provide a larger variety of dynamics, as transients.
This is the reason why the edge of window is strong in our imitation game.
The above speculation suggests the importance of transient chaos,
besides the edge of it, for the adaptation to  a wide range of external
dynamics.

Our imitation process is based on the synchronization of a
player's dynamics to the other's song.  This synchronization
process may not physiologically be unrealistic.  Indeed there has been recent
extensive studies on the entrainment of oscillation.  So far the studies on
synchronization is mostly focused on the visual cortex \cite{synchro},
but there is no reason to suspect the importance of synchronization
in auditory systems.

Evolution of complexity to escape from imitation may play a
keyrole in many fields.
Our imitation game provides one route to the evolution
to complexity.  We may expect that this route can be seen in
some examples in the biological evolution, besides our original
motivation to a bird song.  One is the evolution of a communication code only
within a given group: ( "secret code").
When there are two groups, it is often important to send some messages only
within the same group, so that they are not encoded by the other.
Another example is the Batesian mimicry \cite{Maynard}.  Here again, we have
two groups.  One of the groups can survive better by imitating the
pattern of the other's group, while the other group's advantage in
survival is lost if it is not distinguished well from the other.
Complexity in the patterns of  butterflies in the mimicry
relationship, for example, may be increased through the imitation game
process. Our simulation gives the first conceptual model for the evolution
to complexity by imitation, while modeling with two distinct
species with different roles will be necessary in future for
the study of the above two examples.

\vspace{.2in}

{\em Acknowledgments}:
The authors would like to thank T. Ikegami, K.Tokita, and
S. Adachi for useful discussions. The work is partially supported by
Grant-in-Aids for Scientific
Research from the Ministry of Education, Science, and Culture
of Japan.

\pagebreak

Figure Caption

\vspace{.1in}
Fig.1. Scoretable of the imitation game;
The score of the game between two players with
parameters $a(1)$ and $a(2)$ is plotted with the help
of a gray scale at the corresponding site to $(a(1),a(2))$.
We have computed  the score of
the player 1 with the parameter $a(1)$ against the player with the
parameter $a(2)$, averaged from 100 initial values chosen randomly.
Blank corresponds to 100 wins of the player with $a(2)$,
while the black means 100 wins of the player with $a(1)$.
The darkness of the site $(a(1),a(2))$ increases with
the increase of the winning ratio of the player with $a(1)$.
The parameter$a$ is changed with the increment of 0.04 in the
left table, and 0.003 in the right one.
$T_{imi}=30$, and $T=16$.

\vspace{.1in}

Fig.2 Emergent landscape: Average score for the players with
parameters within $[a_i, a_i +\Delta]$ is plotted
for $a_i=-1+i\times \Delta$, with the bin size $\Delta =0.001$.
Fo figs. 2-4, we have adopted $W=10,L=1$, $T_{imi}=255$,
and $T=32$. For mutational errors random numbers $\delta $ are chosen from
the  Lorenzian  distribution
$P(\delta )= 1 /\{ \mu (1+ (\delta /\mu ) ^2)\} $, to avoid trapping at
intermediate parameters.
Simulation is carried out with the mutation rate $\mu =0.1$,
starting from the initial parameter $a=0.6$ and $\epsilon =.1$.
Sampled for time steps from 1000 to 1500, over all players ( whose number is
fixed at 200).

\vspace{.1in}

Fig.3  Temporal change of the average parameter $a$:
Simulation is carried out with the mutation rate $\mu =0.0005$,
starting from the initial parameter $a=0.6$ and $\epsilon =.1$.
Average of the parameters $a$ over all players are plotted with time.
The total population is fixed at $N=200$.

\vspace{.1in}

Fig.4 Lyapunov exponent versus score;
Simulation is carried out with the mutation rate $\mu =0.001$,
starting from the initial parameter $a=0.9$ and $\epsilon =.1$.
Lyapunov exponents of the song dynamics vs. the score against
all players at the moment are plotted.  Sampled for
time steps from 600 to 800 over all players ( whose number is
fixed at 200), where the parameter value $a$ is concentrated around 1.94.


\begin{thebibliography}{999}
\bibitem{logistic0}
This map is equivalent to the logistic map of the form  $z_{n+1}=bz_n(1-z_n)$,
by the simple transformation of variables;
$ x_n =\frac{4}{b-2} (z_n + 0.5), a=\frac{b(b-2)}{4}$.
See any textbook on chaos for the bifurcation diagram of
the logistic map.
\end{thebibliography}
\end{document}